\documentclass[sigconf]{acmart}
\usepackage{multirow}
\usepackage{caption}

\usepackage{amssymb}
\usepackage{amsmath}
\usepackage{subcaption}
\usepackage{graphicx}
\usepackage{float} 
\usepackage{enumitem}
\usepackage{algorithm} 
\usepackage{algorithmic}
\usepackage{booktabs} 
\usepackage{threeparttable}
\usepackage{makecell}
\usepackage{balance}

\AtBeginDocument{%
  \providecommand\BibTeX{{%
    \normalfont B\kern-0.5em{\scshape i\kern-0.25em b}\kern-0.8em\TeX}}}

\acmConference[KDD '2025]{}{August 03--07,
  2025}{Toronto, ON, Canada}
\acmBooktitle{KDD '2025,
 August 03--07, 2025, Toronto, ON, Canada} 

\begin{document}

\title{LLM4Tag: Automatic Tagging System for  Information Retrieval via Large Language Models}


\newcommand{\chenxu}[1]{{\bf \color{red} [[Chenxu says ``#1'']]}}

\author{Ruiming Tang}
\authornote{Co-first authors with equal contributions.}
\email{tangruiming@huawei.com}
\affiliation{Huawei Noah's Ark Lab\country{China}}

\author{Chenxu Zhu}
\authornotemark[1]
\email{zhuchenxu1@huawei.com}
\affiliation{Huawei Noah's Ark Lab\country{China}}

\author{Bo Chen}
\authornotemark[1]
\email{chenbo116@huawei.com}
\affiliation{Huawei Noah's Ark Lab\country{China}}

\author{Weipeng Zhang}
\email{zhangweipeng12@huawei.com}
\affiliation{Huawei Noah's Ark Lab\country{China}}

\author{Menghui Zhu}
\email{zhumenghui1@huawei.com}
\affiliation{Huawei Noah's Ark Lab\country{China}}

\author{Xinyi Dai}
\email{daixinyi5@huawei.com}
\affiliation{Huawei Noah's Ark Lab\country{China}}

\author{Huifeng Guo}
\email{huifeng.guo@huawei.com}
\affiliation{Huawei Noah's Ark Lab\country{China}}

\renewcommand{\shortauthors}{Ruiming Tang, et al.}

\begin{abstract}

Tagging systems play an essential role in various information retrieval applications such as search engines and recommender systems. Recently, Large Language Models (LLMs) have been applied in tagging systems due to their extensive world knowledge, semantic understanding, and reasoning capabilities. Despite achieving remarkable performance, existing methods still have limitations, including difficulties in retrieving relevant candidate tags comprehensively, challenges in adapting to emerging domain-specific knowledge, and the lack of reliable tag confidence quantification. To address these three limitations above, we propose an automatic tagging system LLM4Tag. First, a graph-based tag recall module is designed to effectively and comprehensively construct a small-scale highly relevant candidate tag set. Subsequently, a knowledge-enhanced tag generation module is employed to generate accurate tags with long-term and short-term knowledge injection. Finally, a tag confidence calibration module is introduced to generate reliable tag confidence scores. Extensive experiments over three large-scale industrial datasets show that LLM4Tag significantly outperforms the state-of-the-art baselines and LLM4Tag has been deployed online for content tagging to serve hundreds of millions of users.

\end{abstract}

\ccsdesc[500]{Information Retrieval~ Tagging Systems}



\keywords{Tagging Systems; Large Language Models; Information Retrieval}



\maketitle

\section{INTRODUCTION}\label{sec:intro}

Tagging is the process of assigning tags, such as keywords or labels, to digital content, products, or users to facilitate organization, retrieval, and analysis. Tags serve as descriptors that summarize key attributes or themes, enabling efficient categorization and searchability, which play a crucial role in information retrieval systems, such as search engines, recommender systems, content management, and social networks~\cite{zhang2011tag,gupta2010survey,bischoff2008can,li2008tag}.
For information retrieval systems, tags are widely used in various stages, including content distribution strategies, ranking algorithms, and operational decision-making processes~\cite{dattolo2010role,ahmadian2022deep}.
Therefore, tagging systems must not only require high accuracy and coverage, but also provide interpretability and strong confidence.

Before the era of Large Language Models (LLMs), the mainstream tagging methods mainly included statistics-based (\textit{i.e.}, TF-IDF-based~\cite{qaiser2018text}, LDA-based~\cite{diaz2010lda}), supervised classification-based (\textit{i.e.}, CNN-based~\cite{zhang2015sensitivity,elnagar2019automatic}, RNN-based~\cite{liu2016recurrent,wang2015unified}), pre-trained model-based methods (\textit{i.e.}, BERT-based~\cite{hasegawa2021bert,bge,ozan2021auto}), 
\textit{etc}.
However, limited by the model capacity, these methods cannot achieve satisfactory results, especially for complex contents. Besides, they heavily rely on annotated training data, resulting in limited generalization and transferability.

The rise of LLMs, with their extensive world knowledge, powerful semantic understanding, and reasoning capabilities, has significantly enhanced the effectiveness of tagging systems. 
LLM4TC~\cite{chae2023large} employs LLMs directly as tagging classifiers and leverages annotated data to fine-tune LLMs. TagGPT~\cite{li2023taggpt} further introduces a match-based recall to filter out a small-scale tag set to address the limited input length of LLMs. ICXML~\cite{zhu2023icxml} proposes an in-context learning algorithm to guide LLMs to further improve performance.

\begin{figure}[h]
\centering
\includegraphics[width=0.98\linewidth]{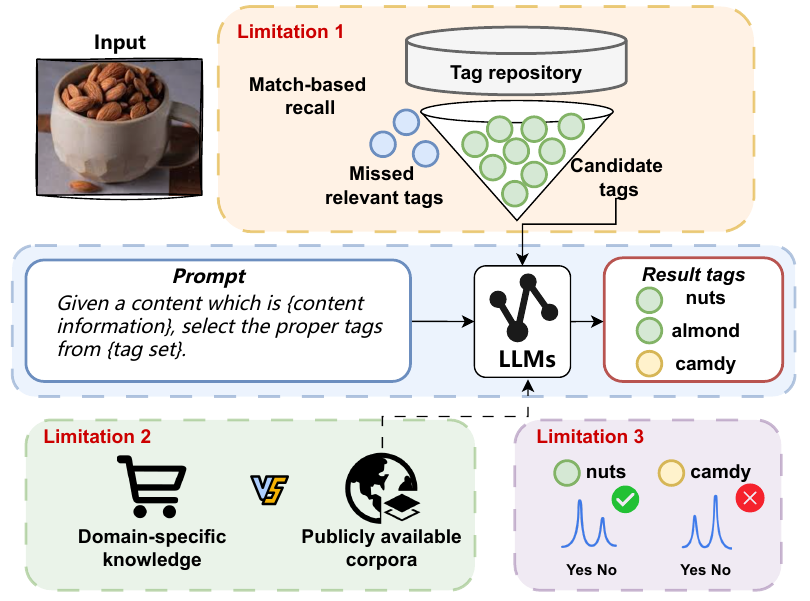}
\caption{LLM-enhanced tagging systems and their three limitations. (L1) Simple match-based recall is prone to missing relevant tags; (L2) The emerging domain-specific knowledge may not align with the pre-trained knowledge of LLMs; (L3) LLMs cannot accurately quantify tag confidence.}
\label{fig:tag_intro}
\end{figure}

However, existing LLM-enhanced tagging algorithms exhibit several critical limitations that require improvement (shown in Figure~\ref{fig:tag_intro}):
\begin{enumerate}[label=(L\arabic*),leftmargin=22pt]
    \item Constrained by the input length and inference efficiency of LLMs, existing methods adopt simple match-based recall to filter out a small-scale candidate tag set~\cite{li2023taggpt,zhu2023icxml}, which is prone to missing relevant tags, thereby reducing accuracy.
    \item General purpose LLMs pre-trained in publicly available corpora exhibit limitations in comprehending emerging domain-specific knowledge within information retrieval, leading to lower accuracy in challenging cases~\cite{chae2023large,li2024ecomgpt}.
    \item Due to the hallucination and uncertainty~\cite{ji2023survey,huang2023look}, LLMs cannot accurately quantify tag confidence, which is crucial for information retrieval applications.
\end{enumerate}

To address the three limitations of existing approaches, we propose an automatic tagging system called LLM4Tag, which consists of three key modules.
Specifically, to improve the completeness of candidate tags (L1), we propose a graph-based tag recall module designed to construct small-scale, highly relevant candidate tags from a massive tag repository efficiently and comprehensively.
To enhance domain-specific knowledge and adaptability to emerging information of general-purpose LLMs (L2), a knowledge-enhanced tag generation module that integrates long-term supervised knowledge injection and short-term retrieved knowledge injection is designed to generate accurate tags.
Moreover, a tag confidence calibration module is introduced to generates reliable tag confidence scores, ensuring more robust and trustworthy results (L3). 


To summarize, the main contributions of this paper can be highlighted as follows:
\begin{itemize}[leftmargin=12pt]
\item We propose an LLM-enhanced tagging framework LLM4Tag, characterized by completeness, continuous knowledge evolution, and quantifiability.
\item  To address the limitations of existing approaches, LLM4Tag integrates three key modules: graph-based tag recall, knowledge-enhanced tag generation, and tag confidence calibration, ensuring the generation of accurate and reliable tags.
\item LLM4Tag achieves state-of-the-art in three large-scale industrial datasets with detailed analysis that provides a deeper understanding of model performance.
Moreover, LLM4Tag has been deployed online for content tagging, serving hundreds of millions of users. 
\end{itemize}




\begin{figure*}[t]
  \centering
  \includegraphics[width=1.0\linewidth]{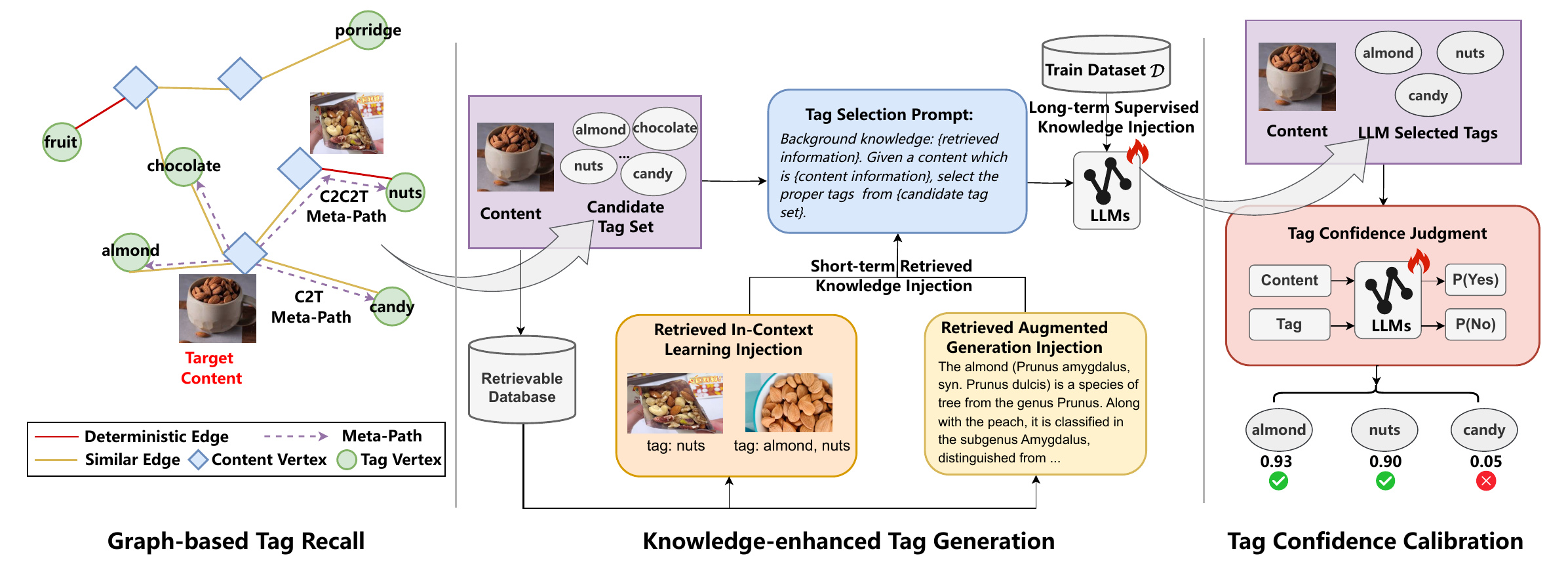}
  \caption{The overall framework of LLM4Tag architecture of LLM4Tag, consisting of three modules: graph-based tag recall module, knowledge-enhanced tag generation module, and tag confidence calibration module.}
  \label{fig: overall}
\end{figure*}

\section{RELATED WORK}

In this section, we briefly review traditional tagging systems and LLM-enhanced tagging systems.

\subsection{Traditional Tagging Systems}
Traditional tagging systems~\cite{gupta2010survey,mishne2006autotag,choi2016automatic} generally employ multi-label classification models, which utilize human-annotated tags as ground truth labels and employ content descriptions as input to predict. Qaiser et al.~\cite{qaiser2018text} utilize TF-IDF to categorize the tag while Diaz et al.~\cite{diaz2010lda} employ LDA to automatically tag the resource based on the most likely tags derived from the latent topics identified. The advent of deep learning has led to the proposal of RNN-based~\cite{liu2016recurrent} and CNN-based~\cite{zhang2015sensitivity} methods for achieving multi-label learning, which are directly applied to tagging systems~\cite{wang2015unified, elnagar2019automatic}. Hasegawa et al.~\cite{hasegawa2021bert} further adopt the BERT pre-training technique in their tagging systems. Recently, with the growing popularity of pre-trained Small Language Models (SLMs), numerous pre-training embedding models, such as BGE~\cite{bge}, GTE~\cite{gte}, and CONAN~\cite{conan}, have been proposed and directly employed in tagging systems through domain knowledge fine-tuning.

Nonetheless, the capabilities of these models are constrained by their limited model capacity, particularly in the presence of complex content. Additionally, they depend excessively on annotated training data, resulting in sub-optimal generalization and transferability.

\subsection{LLM-Enhanced Tagging Systems}
With Large Language Models (LLMs) achieving remarkable breakthroughs in natural language processing~\cite{achiam2023gpt,touvron2023llama,guo2025deepseek,brown2020language} and information retrieval systems~\cite{zhu2023large,lin2023can}, LLM-enhanced tagging systems have received much attention and have been actively explored currently~\cite{wang2023large,sun2023text,chae2023large,li2023taggpt,zhu2023icxml}.
Wang et al.~\cite{wang2023large} employ LLMs as a direct tagging classifier, while Sun et al.~\cite{sun2023text} introduce clue and reasoning prompts to further enhance performance. LLM4TC~\cite{chae2023large} undertakes studies on diverse LLMs architectures and leverages annotated samples to fine-tune the LLMs. TagGPT~\cite{li2023taggpt} introduces an early match-based recall mechanism to generate candidate tags from a large-scale tag 
repository with textual clues from multimodal data. ICXML~\cite{zhu2023icxml} proposes a two-stage framework through in-context learning to guide LLMs to align with the tag space.

However, the aforementioned works suffer from three critical limitations (mentioned in Section 1): (1) difficulties in comprehensively retrieving relevant candidate tags, (2) challenges in adapting to emerging domain-specific knowledge, and (3) the lack of reliable tag confidence quantification. 
To this end, we propose LLM4Tag, an automatic tagging system, to address the aforementioned limitations.


\section{Methodology}
In this section, we present our proposed LLM4Tag framework in detail. We start by providing an overview of the proposed framework and then give detailed descriptions of the three modules in LLM4Tag.

\subsection{Overview}
As illustrated in Figure~\ref{fig: overall}, our proposed LLM4Tag framework consists of three major modules: (1) Graph-based Tag Recall, (2) Knowledge-enhanced Tag Generation, and (3) Tag Confidence Calibration, which respectively provides completeness, continual knowledge evolution, and quantifiability.

\textbf{Graph-based Tag Recall} module is responsible for retrieving a small-scale, highly relevant candidate tag set from a massive tag repository. Based on a scalable content-tag graph constructed dynamically, graph-based tag recall is utilized to fetch dozens of relevant tags for each content. 

\textbf{Knowledge-enhanced Tag Generation} module is designed to accurately generate tags for each content via Large Language Models (LLMs). To address the lack of domain-specific and emerging knowledge in general-purpose LLMs, this module implements a scheme integrating the injection of both long-term and short-term domain knowledge, thereby achieving continual knowledge evolution.

\textbf{Tag Confidence Calibration} module is aimed to generate a quantifiable and reliable confidence score for each tag, thus alleviating the issues of hallucination and uncertainty in LLMs. Furthermore, the confidence score can be employed as a relevance metric for downstream information retrieval tasks.


\subsection{Graph-based Tag Recall}
Given the considerable magnitude of tags (millions) in industrial information retrieval system, the direct integration of the whole tag repository into LLMs is impractical due to the constrained nature of the context window and inference efficiency of LLMs. 
Existing approaches~\cite{li2023taggpt,zhu2023icxml} adopt simple match-based tag recall to filter out a small-scale candidate tag set based on small language models (SLMs), such as BGE~\cite{bge}. However, they are prone to missing relevant tags due to the limited capabilities of SLMs. 
To address this issue and improve the comprehensiveness of the retrieved candidate tags, we construct a semantic graph globally and propose a graph-based tag recall module.

Firstly, we initial an undirected graph $\mathcal{G}$ with contents and tags as:
\begin{equation}
\begin{aligned}
\mathcal{G}=\{\mathcal{V},\mathcal{E}\}~,
\end{aligned}
\end{equation}
where vertex set $\mathcal{V}=\{\mathcal{C},\mathcal{T}\}$ is the set of existing content vertices $\mathcal{C}$ and all tag vertices $\mathcal{T}$. As for the edge set $\mathcal{E}$, it contains two types of edges, called \textit{Deterministic Edges} and \textit{Similarity Edges}.

\textbf{Deterministic Edges} only connect between content vertex $c$ and tag vertex $t$, formulated as $e_{c-t}^d$, which indicates that content $c$ is labeled with tag $t$ based on historical annotation data.
To ease the high sparsity of the deterministic edges in the graph $\mathcal{G}$, we further introduce semantic similarity-based edges (\textbf{Similarity Edges}) that connect not only between content vertex $c$ and tag vertex $t$, but also between different content vertices, formulated as $e_{c-t}^s$ and $e_{c-c}^s$, respectively. 

Specifically, for the $i$-th vertex $v^i \in \mathcal{V}$ in graph $\mathcal{G}$, we summarize all textual information (\textit{i.e.}, title and category in content, tag description) as $text^{i}$ and vectorize it with an encoder to get a semantic representation $\boldsymbol{r}^i$:
\begin{equation}
\begin{aligned}
    \textit{\textbf{r}}^i &= \operatorname{Encoder}(text^{i})~,
\end{aligned}
\end{equation}
where $\operatorname{Encoder}$ is a small language model, such as BGE~\cite{bge}. Then the similarity distance of two different vertices $v^i, v^j$ can be computed as:
\begin{equation}
\begin{aligned}
    \operatorname{Dis}(v^i, v^j) &= \frac{\textit{\textbf{r}}^i\cdot\boldsymbol{r}^j}{\lVert\boldsymbol{r}^i\rVert \lVert\boldsymbol{r}^j\rVert}~.
\end{aligned}
\end{equation}

After obtaining the similarity estimations, we can use a threshold-based method to determine the similarity edge construction, \textit{i.e.},
\begin{itemize}[leftmargin=12pt]
    \item $e_{c-t}^s$ connects the content $c$ and the tag $t$ if the similarity distance between them exceeds $\delta_{c-t}$.
    \item $e_{c-c}^s$ connects the similar contents when their similar distance exceeds $\delta_{c-c}$.
\end{itemize}

In this way, we can construct a basic content-tag graph with deterministic/similarity edges. Then, when a new content $c$ appears that needs to be tagged, we dynamically insert it into this graph by adding similarity edges.
Next, we define two types of meta-paths (\textit{i.e.}, C2T meta-path and C2C2T meta-path) and adopt the meta-path-based approach to recall candidate tags.

\textbf{C2T Meta-Path}: Based on the given content $c$, we first recall the tags which is connected directly to $c$ as the candidate tags. The meta-path can be defined as:
\begin{equation}
\begin{aligned}
    p^{C2T} = c\overset{s}{\rightarrow} t~, 
\end{aligned}
\end{equation}
where $\overset{s}{\rightarrow}$ is the  similarity edge.

\textbf{C2C2T Meta-Path}: C2C2T contains two sub-procedures: C2C and C2T. C2C is aimed at discovering similar contents while C2T further attempts to recall the deterministic tags from these similar contents. The meta-path can be formulated as:
\begin{equation}
\begin{aligned}
    p^{C2C2T} = c\overset{s}{\rightarrow}c\overset{d}{\rightarrow}t~,
\end{aligned}
\end{equation}
where $\overset{d}{\rightarrow}$ is the deterministic edge and $\overset{s}{\rightarrow}$ is the similarity edge.

With these two types of meta-paths, we can generate a more comprehensive candidate tag set for content $c$ as
\begin{equation}
\begin{aligned}
    \Phi(c) &=  \Phi^{C2T}\left(c\right) \cup \Phi^{C2C2T}\left(c\right)~,
\end{aligned}
\end{equation}
where $\Phi^{C2T}\left(c\right)$ is retrieved by C2T meta-path and $\Phi^{C2C2T}$ is retrieved by C2C2T meta-path.
Notably, the final tagging results of LLM4Tag for the content $c$ will also be added to the graph as deterministic edges, enabling dynamic scalability of the graph.

Compared to simple match-based tag recall, our graph-based tag recall leverages semantic similarity to construct a global content-tag graph and incorporates a meta-path-based multi-hop recall mechanism to enhance candidate tags completeness, which will be demonstrated in Sec~\ref{sec:exp_retrieval}. 


\subsection{Knowledge-enhanced Tag Generation}
After obtaining the candidate tag set, we can directly use the Large Language Models (LLMs) to select the most appropriate tags. However, due to the diversity and industry-specific nature of the information retrieval system applications, domain-specific knowledge varies significantly across different scenarios. That is, the same content and tags may have distinct definitions and interpretations depending on the specific application context.
Furthermore, the domain-specific knowledge is emerged continually and constantly at an expeditious pace. As a result, the general-purpose LLMs have difficulty in understanding the emerging domain-specific information, such as newly listed products, emerging hot news, or newly added tags, leading to a lower accuracy on challenging cases.

To address the lack of emerging domain-specific information in LLMs, we devise a knowledge-enhanced tag generation scheme that takes into account both long-term and short-term domain-specific knowledge by two key components, namely \textit{Long-term Supervised Knowledge Injection} (LSKI), \textit{Short-term Retrieved Knowledge Injection} (SRKI).


\begin{figure}[h]
    \centering
    \includegraphics[width=0.48\textwidth]{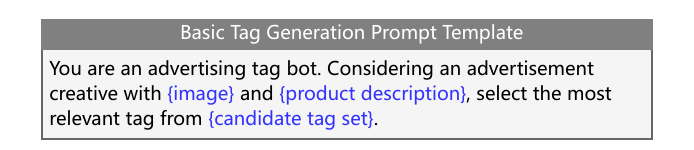}
    \caption{Prompt template for basic tag generation in advertisement creatives tagging scenario.}
    \label{fig:prompt1}
\end{figure}

\subsubsection{Long-term Supervised Knowledge Injection.} For long-term domain-specific knowledge, we first construct a training dataset $\mathcal{D}$ and adopt a basic prompt template as ${Template_b}$ for tag generation (shown in Figure~\ref{fig:prompt1}). 
\begin{equation}
\begin{aligned}
    \mathcal{D} &= \{(x_i, y_i)\}_{i=1}^N~,\\
    x_i &= {Template_b}(c_i, \Phi(c_i))~,
\end{aligned}
\end{equation}
where $N$ is the size of training dataset. Notably, to ensure the comprehensiveness of domain-specific knowledge, we employ the principle of diversity for sample selection and obtain correct answers $y_i$ by combining LLMs generation with human expert annotations.

After obtaining the training set, we leverage the causal language modeling objective for LLM Supervised Fine-Tuning (SFT):
\begin{equation}
    \max_{\Theta}\sum_{i=1}^N\, \sum_{j=1}^{|y_i|}\log P_{\Theta}\left(y_{i,j} \mid x_i,y_{i,<j}\right)~,
    \label{eq:loss}
\end{equation}
where $\Theta$ is the parameter of LLM, $y_{i,j}$ is the $j$-th token of the textual output $y_{i}$, and $y_{i,<j}$ denotes the tokens before $y_{i,j}$ in the $i$-th samples.

By adopting this approach, we can effectively integrate the domain-specific knowledge from information retrieval systems into LLMs, thus improving the tagging performance.

\subsubsection{Short-term Retrieved Knowledge Injection.} Although LSKI effectively provides domain-specific knowledge, continuously incorporating short-term knowledge through LLM fine-tuning is highly resource-intensive, especially given the rapid emergence of new domain knowledge.
Additionally, this approach suffers from poor timeliness, making it more challenging to adapt to rapidly evolving content in information retrieval systems, particularly for emerging hot topics.

Therefore, we further introduce a short-term retrieved knowledge injection (SRKI). 
Specifically, we derive two retrieved knowledge injection methods: \textit{retrieved in-context learning injection} and \textit{retrieved augmented generation injection}.




\textbf{Retrieved In-Context Learning Injection}.
We first construct a retrievable sample knowledge base (including contents and their correct/incorrect annotated tags) and continuously append newly emerging samples. 
Then, given the target content $c$, this composition retrieves $n$ relevant samples from the sample knowledge base. 
This approach not only leverages the few-shot in-context learning capability of LLMs but also enables them to quickly adapt to emerging domain knowledge, enhancing tagging accuracy for challenging cases.


\textbf{Retrieved Augmented Generation Injection}. 
Given the content $c$ and the candidate tag set $\Phi(c)$, this composition retrieves relevant descriptive corpus from web search and domain knowledge base. It can retrieve extensive information that assists LLMs in understanding unknown domain-specific knowledge or new knowledge, such as the definition of terminology in the content/tag or some manually defined tagging rules.

\begin{figure}[h]
    \centering
    \includegraphics[width=0.48\textwidth]{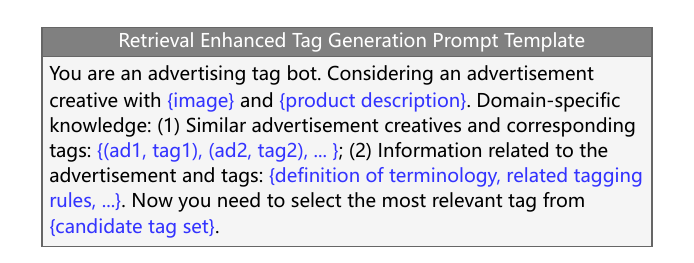}
    \caption{Prompt template for retrieval enhanced tag generation in advertisement creatives tagging scenario.}
    \label{fig:prompt2}
\end{figure}

After obtaining the retrieved knowledge, we design a prompt template, ${Template_r}$,  (shown in Figure~\ref{fig:prompt2}) to integrate knowledge with the content $c$ and candidate tag set $\Phi(c)$ to provide the in-context guidance for LLMs to predict the most appropriate tags for content $c$ as:
\begin{equation}
\begin{aligned}
      \Gamma(c)  &=\operatorname{LLM}({Template_r}(c, \Phi(c), R(c)))~, \\
      &= \left\{t^c_1, t^c_2, \cdots, t^c_m\right\}~,
\end{aligned}
\end{equation}
where $R(c)$ is the retrieved knowledge above, and $m$ is the number of appropriate tags generated by LLMs.

\subsection{Tag Confidence Calibration}
After tag generation, there still exist two serious problems for real-world applications: (1) the hallucination due to the uncertainty of LLMs, which leads to generating irrelevant or wrong tags; (2) the necessity of assigning a quantifiable relevance score for each tag for the sake of downstream usage in the information retrieval systems (\textit{e.g.}, recall and marketing). 


\begin{figure}[h]
    \centering
    \includegraphics[width=0.48\textwidth]{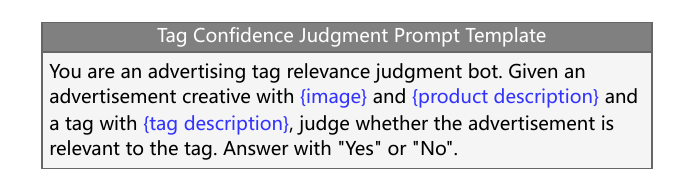}
    \caption{Prompt template for tag confidence judgment in advertisement creatives tagging scenario.}
    \label{fig:prompt3}
\end{figure}

To handle these two problems, the tag confidence calibration module is adopted. Specifically, given a target content $c$ and a certain tag $t^c \in \Gamma(c)$, we derive a prompt template, ${Template_c}$ (shown in Figure~\ref{fig:prompt3}), to leverage the reasoning ability of LLMs to achieve a tag confidence judgment task, \textit{i.e.}, whether $c$ and $t^c$ is relevant. Then we extract the probability of the token in the LLM result to get a confidence score $\operatorname{Conf}(c, t^c)$:
\begin{equation}
\begin{aligned}
    \boldsymbol{s} &= \operatorname{LLM}({Template_c}(c, t^c))\,\in\mathbb{R}^{V}, \\
    \operatorname{Conf}(c, t^c) &= \frac{\exp\left(\boldsymbol{s}\left[``Yes"\right]\right)}{\exp\left(\boldsymbol{s}\left[``Yes"\right]\right)+\exp\left(\boldsymbol{s}\left[``No"\right])\right)} \,\in(0,1)~,
\end{aligned}
\end{equation} 
where $\boldsymbol{s}$ is the probability score for all tokens, and $V$ is the vocabulary size of LLMs.

After obtaining the confidence score $\operatorname{Conf}(c, t)$, we implement self-calibration for the results by eliminating those tags with low confidence, achieving a better performance by mitigating the hallucination problem. Furthermore, this confidence score can be directly set as a relevance metric for the downstream tasks. 

\textbf{Tag Confidence Training.}
In order to make the confidence score more consistent with the requirements of information retrieval, we construct a confidence training dataset $\mathcal{D}'$ as:
\begin{equation}
\begin{aligned}
    \mathcal{D}' &= \{(x_i', y_i')\}_{i=1}^M,\\
    x_i' &= \operatorname{Prompt_c}(c_i, t_i)~,\\
    y_i' &\in \{``Yes", ``No"\}~,
\end{aligned}
\end{equation}
where $y_i$ is annotated by experts and $M$ is the size of training dataset. Then we leverage the causal language modeling objective, which is the same as Equation~(\ref{eq:loss}) to perform supervised fine-tuning. In that case, the confidence score predicted by this module aligns with the requirements of the information retrieval systems, thereby facilitating the calibration of incorrect tags.





\begin{table*}
\caption{Performance comparison of different methods. Note that different tasks, multi-tag tasks (Brower News) and single-tag tasks (Advertisement Creatives and Search Query), have different metrics. The best result is given in bold, and the second-best value is underlined. "RI" indicates the relative improvements of LLM4Tag over the corresponding baseline.}
\label{table:all}
 \renewcommand\arraystretch{1.1}
\begin{threeparttable}[htbp]{
\begin{tabular}{c|cccc|cccc|cccc}
\midrule
 {\multirow{2}{*}{{Model}}} & \multicolumn{4}{c|}{{Browser News}} & \multicolumn{4}{c|}{{Advertisement Creatives}} & \multicolumn{4}{c}{{Search Query}} \\ 
\cmidrule(lr){2-5}\cmidrule(lr){6-9}\cmidrule(lr){10-13}
  & Acc@1 & Acc@2 & Acc@3 & RI & Precision & Recall & F1 & RI & Precision & Recall & F1 & RI \\ 
\midrule
  BGE & 0.7427 & 0.6584 & 0.5976 & 29.8\% & 0.7817 & 0.7396   & 0.7601 & 18.5\% & 0.6364 & 0.5122  & 0.5676&  56.2\% \\
  GTE & 0.7292 & 0.6507 & 0.5941 & 31.3\% & 0.7369 & 0.7026  & 0.7194& 25.0\%  & 0.6129 & 0.4634 & 0.5278 & 67.9\% \\
 CONAN  & 0.7568 & 0.6814 & 0.6266 & 25.5\%  & 0.7491  & 0.7194 & 0.7339& 22.6\% & 0.6056 & 0.5244 & 0.5621& 57.9\% \\
\midrule
 TagGPT & 0.8351 & 0.7813 & 0.7424 &   9.5\%  & 0.8454 & 0.7997  & 0.8219& 9.4\% & 0.8421 & 0.7805  & 0.8101 & 9.7\% \\
 ICXML & 0.8398 & 0.7883 & 0.7560 & 8.4\%  & 0.8492  & 0.8025  & 0.8252 & 9.0\% & 0.8600 & 0.7840 & 0.8202& 8.3\% \\
 LLM4TC & \underline{0.8602} & \underline{0.8069} & \underline{0.8235} & 3.7\% & \underline{0.8726} & \underline{0.8245}  & \underline{0.8479} & 6.1\% & \underline{0.9028} & \underline{0.8025} & \underline{0.8497} & 4.5\% \\
\midrule
 \textbf{LLM4Tag} & \textbf{0.9041}  & \textbf{0.8511} & \textbf{0.8273} & - & \textbf{0.9138} & \textbf{0.8857} & \textbf{0.8995} &-  & \textbf{0.9325} & \textbf{0.8485} & \textbf{0.8885} & -\\  
  \midrule          
\end{tabular}
}
\end{threeparttable}
\end{table*}

\section{EXPERIMENTS}\label{sec:exp}
In this section, we conduct extensive experiments to answer the following research questions:
\begin{itemize}
    \item[\textbf{RQ1}] How does LLM4Tag perform in comparison to existing tagging algorithms?
    \item[\textbf{RQ2}] How effective is the graph-based tag recall module?
    \item[\textbf{RQ3}] Does the injection of domain-specific knowledge enhance the tagging performance?
    \item[\textbf{RQ4}] What is the impact of the tag confidence calibration module?
\end{itemize}

\subsection{Experimental Settings}
\subsubsection{Dataset}
We conducted experiments on a mainstream information distribution platform with hundreds of millions of users and sampled three representative industrial datasets  from online logs to ensure consistency in data distribution, containing two types of tasks: (1) multi-tag task (\textit{Browser News}), and (2) single-tag task (\textit{Advertisement Creatives} and \textit{Search Query}).

\begin{itemize}[leftmargin = 12 pt]
\item \textbf{Browser News} dataset includes popular news articles and user-generated videos, primarily in the form of text, images, and short videos. This is a multi-tag task, wherein the objective is to select multiple appropriate tags for each content from a massive tag repository (more than 100,000 tags). Around 30,000 contents are randomly collected as the testing dataset through expert annotations.

\item \textbf{Advertisement Creatives} dataset includes ad creatives, including cover images, copywriting, and product descriptions from advertisers.
The task for this dataset is a single-tag task, where we need to select the most relevant tag to the advertisement from a well-designed tag repository (more than 1,000 tags) and we collect around 10,000 advertisements randomly as the testing dataset through expert annotation.
\item \textbf{Search Query} dataset primarily consists of user search queries from a web search engine, used for user intent classification.
The task for this dataset is also a single-tag task, where the most probable intent needs to be selected as the tag for each query. The size of the tag repository is about 1,000, and 2,000 queries are collected and manually tagged as the testing dataset.
 \end{itemize}
 
 \subsubsection{Baselines}
To evaluate the superiority and effectiveness of our proposed model, we compare LLM4Tag with two classes of existing models: 
\begin{itemize}[leftmargin = 12 pt]
\item \textbf{Traditional Methods} that encode the contents and tags by leveraging pre-trained language models and select the most relevant tags according to cosine distance for each content as the result. Here we compare three different pre-trained language models.
\textbf{BGE}~\cite{bge} pre-trains the models with retromae on large-scale pairs data using contrastive learning. \textbf{GTE}~\cite{gte} further proposes multi-stage contrastive learning to train the text embedding. \textbf{CONAN}~\cite{conan} maximizes the utilization of more and higher-quality negative examples to pre-train the model.
\item \textbf{LLM-Enhanced Methods} that utilize large language models to assist the tag generation. \textbf{TagGPT}~\cite{li2023taggpt} proposes a zero-shot automated tag extraction system through prompt engineering via LLMs. \textbf{ICXML}~\cite{zhu2023icxml} introduces a two-stage tag generation framework, involving generation-based label shortlisting and label reranking through in-context learning. \textbf{LLM4TC}~\cite{chae2023large} further leverages fine-tuning using domain knowledge to improve the performance of tag generation.
 \end{itemize}

\subsubsection{Evaluation Metrics}
For multi-tag tasks, due to the excessive number of tags (millions), we can not annotate all the correct tags and thus only directly judge whether the results generated by the model are correct or not. In this case, we define Acc@k to evaluate the performance:
\begin{equation}
\begin{aligned}
\operatorname{Acc@k} &=\frac{1}{N'}\sum_{i=1}^{N'}\sum_{j=1}^{k'}  \frac{\mathbb{I}\,{\left(T_i[j]\right)}}{k'}~, \\
k' &= \min(k, len(T_i))~,\\
\mathbb{I}\,{\left(T_i[j]\right)}&= \begin{cases}
1, & \text{$T_i[j]$ is right}~,\\
0, &  \text{otherwise}~,
\end{cases}
\end{aligned}
\end{equation}
where $T_i[j]$ is the $j$-th generated tag of the $i$-th content and $N'$ is the size of test dataset. It is worth noticing there exists contents that do not have $k$ proper tags, thus we allow the number of generated tags to be less than $k$. 

For the single-tag task, we adopt Precision, Recall, and F1 following previous works~\cite{li2023taggpt,chae2023large}. Higher values of these metrics indicate better performance.

Moreover, we report Relative Improvement (RI) to represent the relative improvement our model achieves over the compared models. Here we calculate the average RI of the above all metrics.

\subsubsection{Implementation Details} 
In the selection of LLMs, we select Huawei's large languge model PanGu-7B~\cite{zeng2021pangu,wang2023pangu}.
For the graph-based tag recall module, we choose BGE~\cite{bge} as the encoder model. $\delta_{c-t}$ and $\delta_{c-c}$ are set as $0.5$ and $0.8$, respectively. Besides, we set maximum recall numbers for different meta-paths, $15$ for C2T meta-path and $5$ for C2C2T meta-path.
For knowledge-enhanced tag generation module, the size of the training dataset in long-term supervised knowledge injection contains approximately $10,000$ annotated samples and the tuning is performed every two weeks. As for the short-term retrieved knowledge injection, the retrievable database is updated in real-time and we retrieve at most $3$ relevant samples/segments for in-context learning injection and augmented generation injection, respectively.
For the tag confidence calibration module, we eliminate tags with confidence scores less than $0.5$ and rank the remaining tags in order of confidence scores as the result.




\subsection{Result Comparison \& Deployment (RQ1)}
Table~\ref{table:all}  summarizes the performance of different methods on three industrial datasets, from which we have the following observations:

\begin{itemize}[leftmargin = 12 pt]
    \item \textbf{Leveraging large language models (LLMs) brings benefit to model performance.} TagGPT, ICXML, and LLM4TC, utilize LLMs to assist the tag generation, achieving better performances than other small language models (SLMs), such as BGE, GTE, and CONAN.
    This phenomenon indicates that the world knowledge and reasoning capabilities of LLMs enable better content understanding and tag generation, significantly improving tagging effectiveness.

    \item \textbf{Introducing domain knowledge can significantly improve performance.} Although LLMs benefit from general world knowledge, there remains a significant gap compared with domain-specific knowledge. Therefore, LLM4TC injects domain knowledge by fine-tuning the LLMs and achieves better performance than other baselines in all metrics, which validates the importance of domain knowledge injection.
    \item \textbf{The superior performance of LLM4Tag.} We can observe from Table~\ref{table:all} that LLM4Tag yields the best performance on all datasets consistently and significantly, validating the superior effectiveness of our proposed LLM4Tag. Concretely, LLM4Tag beats the best baseline by \textbf{3.7\%}, \textbf{6.1\%}, and \textbf{4.5\%} on three datasets, respectively. This performance improvement is attributed to the advanced nature of our LLM4Tag, including more comprehensive graph-based tag recall, deeper domain-specific knowledge injection, and more reliable confidence calibration.
    \item \textbf{Notably, LLM4Tag has been deployed online and covers all the traffic.} We randomly resampled the online data, and the online report shows consistency between the improvements in the online metrics and those observed in the offline evaluation. Now, LLM4Tag has been deployed in the content tagging system of these three online applications, serving hundreds of millions of users daily.

\end{itemize}


\subsection{The Effectiveness of Graph-based Tag Recall Module (RQ2)}
\label{sec:exp_retrieval}
In this subsection, we compare our proposed graph-based tag recall module with match-based recall to validate the effectiveness of candidate tags retrieval over the Browser News Dataset. For fairness, both methods use the same pre-trained language model BGE to encode contents and tags, and the number of candidate tags is fixed as 20.
We define two metrics to evaluate the performance: \textit{\#Right} means the average number of correct tags in candidate tags, and \textit{HR\#k} means the proportion of cases where at least $k$ correct tags are hit in the candidate tag set.

\begin{table}[htbp]
	\caption{Performance comparison between different recall types over the Browser News Dataset.}
	\label{tab:recall}
	\centering
	\resizebox{0.4\textwidth}{!}{
	 \renewcommand\arraystretch{1.1}
		\begin{tabular}{c|cccc}
			\midrule
            Recall Type &  \#Right &  HR\#1 & HR\#2 & HR\#3   \\ \midrule
            Match-based &  4.48 & 0.9586 & 0.8841 & 0.7643 \\
            Ours & 5.37 & 0.9745 & 0.9212 & 0.8425  \\ 
			\midrule
		\end{tabular}
		} 
\end{table}

As shown in Table~\ref{tab:recall}, we can find that our graph-based recall method can significantly improve the quality of candidate tags. The metrics \#Right and  HR\#3 increase by $19.8\%$ and $10.2\%$ respectively, which demonstrates that our method yields a more complete and comprehensive candidate tag set via a meta-path-based multi-hop recall mechanism. Moreover, the lifting of HR\#1 illustrates that our method can recall the correct tags when the match-based method encounters challenges in hard cases and fails to select the relevant tags.

\begin{figure}[h]
  \centering
  \vspace{-1.0em}
  \includegraphics[width=0.98\linewidth]{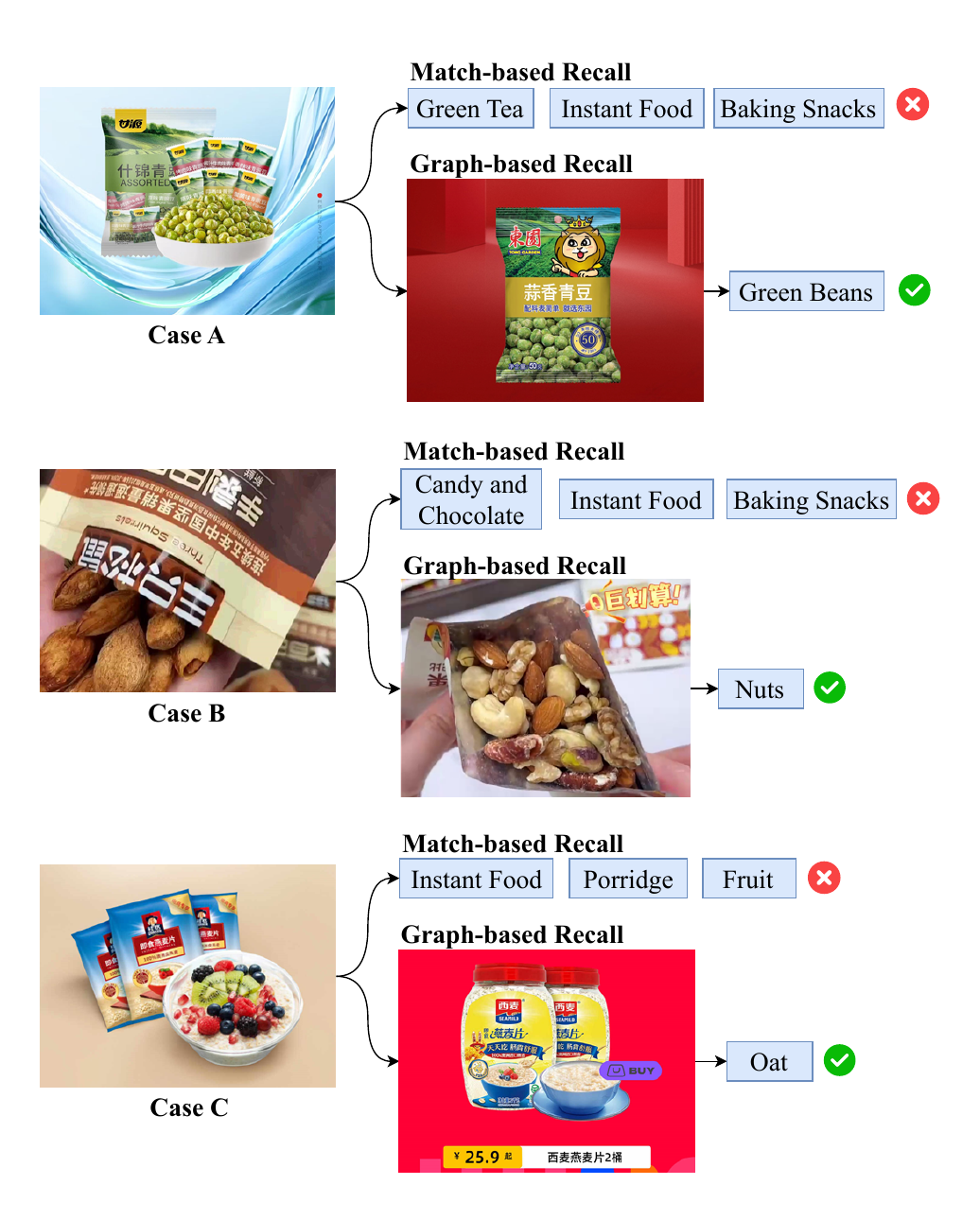}
  \vspace{-1.0em}
  \caption{The online cases for verifying the effectiveness of graph-based tag recall.}
  \label{fig:case_retrieval}
\end{figure}

Besides, to verify the effectiveness and interpretability of our proposed graph-based tag recall, we randomly select some cases in our deployed tagging scenario and visualize the recall results in Figure~\ref{fig:case_retrieval}.
It can be observed that, when match-based recall fails to select the correct tags for some challenging cases, our method effectively retrieves accurate tags by C2C2T meta-path multi-hop traversal in the graph, thus avoiding missing correct tags due to the limited capabilities of SLMs.



\subsection{The Effectiveness of Knowledge-enhanced Tag Generation (RQ3)}
In order to systematically evaluate the contribution of domain knowledge-enhanced tag generation (KETG) in our framework, we have designed the following variants:

\begin{itemize}[leftmargin = 12 pt]
\item \textbf{LLM4Tag (w/o KETG)} removes both long-term supervised knowledge injection (LSKI) and short-term retrieved knowledge injection (SRKI), and selects tags using native LLMs.
\item \textbf{LLM4Tag (w/o LSKI)} removes LSKI and only maintains SRKI to inject the short-term domain-specific knowledge.
\item \textbf{LLM4Tag (w/o SRKI)} removes SRKI and only maintains LSKI to inject the long-term domain-specific knowledge.
\item \textbf{LLM4Tag (Ours)} incorporates both LSKI and SRKI to inject the long/short-term domain-specific knowledge.
\end{itemize}

\begin{figure}[h]
    \centering
    \includegraphics[width=0.36\textwidth]{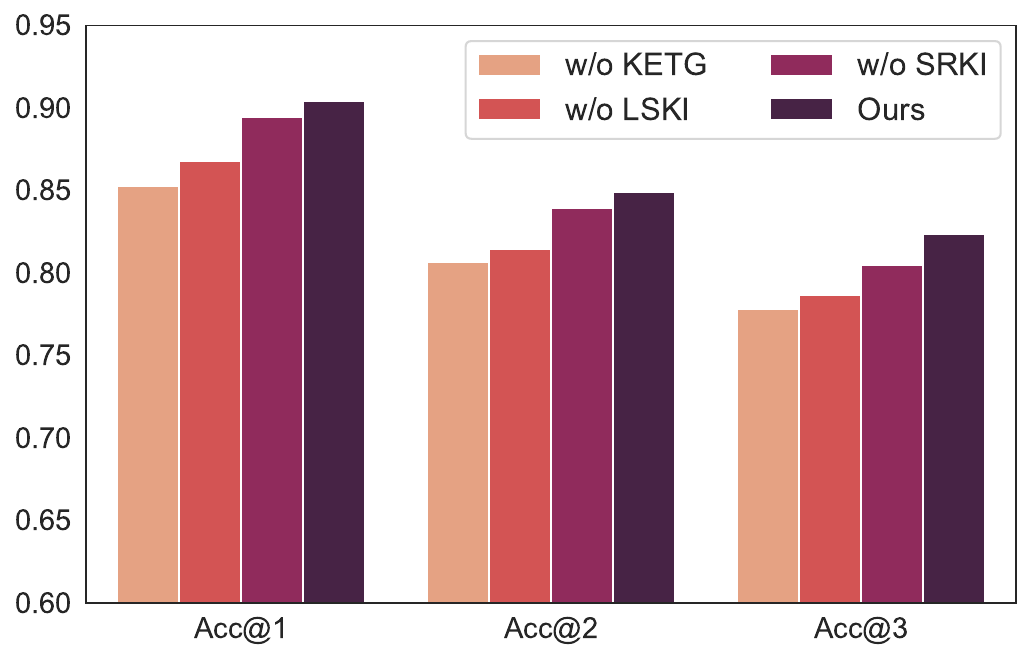}
    \caption{Ablation study about the effectiveness of knowledge-enhanced tag generation module in LLM4Tag.}
    \label{fig:dke}
\end{figure}
Figure~\ref{fig:dke} presents the comparative results on Browser News Dataset, revealing three key findings:

\begin{itemize}[leftmargin = 12 pt]
\item The complete framework achieves optimal performance, demonstrating the synergistic value of combining supervised fine-tuning in long-term supervised knowledge injection with non-parametric short-term retrieved knowledge injection.
\item The removal of either component in knowledge-enhanced tag generation module causes measurable degradation.
Among them, the removal of long-term knowledge results in a greater decline, indicating that long-term knowledge may cover a broader range of domain-specific knowledge and highlighting the importance of SFT in model knowledge injection.

\item The most basic variant (w/o KETG) exhibits the lowest performance, highlighting the crucial role of domain adaptation in specialized tagging tasks within information retrieval systems.
\end{itemize}



\subsection{The Effectiveness of Tag Confidence Calibration (RQ4)}
To validate the effectiveness of the tag confidence calibration module, we evaluate model performance on the Browser News Dataset and use different confidence thresholds to achieve different pruning rates. Here we define a metric \textit{Coverage@k} to evaluate the cover rate of final results as:
\begin{equation}
\begin{aligned}
\text{Coverage}@k &= \frac{1}{N'} \sum_{i=1}^{N'} \mathbb{I}\,{\left(|T_i| \geq k\right)}~,\\
\mathbb{I}\,{\left(|T_i| \geq k\right)}&= \begin{cases}
1, & |T_i| \geq k~,\\
0, &  \text{otherwise}~,
\end{cases}
\end{aligned}
\end{equation}
where $T_i$ is the result tags of the $i$-th content and $N'$ is the size of testing dataset. 


\begin{figure}[h]
\centering
\includegraphics[width=0.38\textwidth]{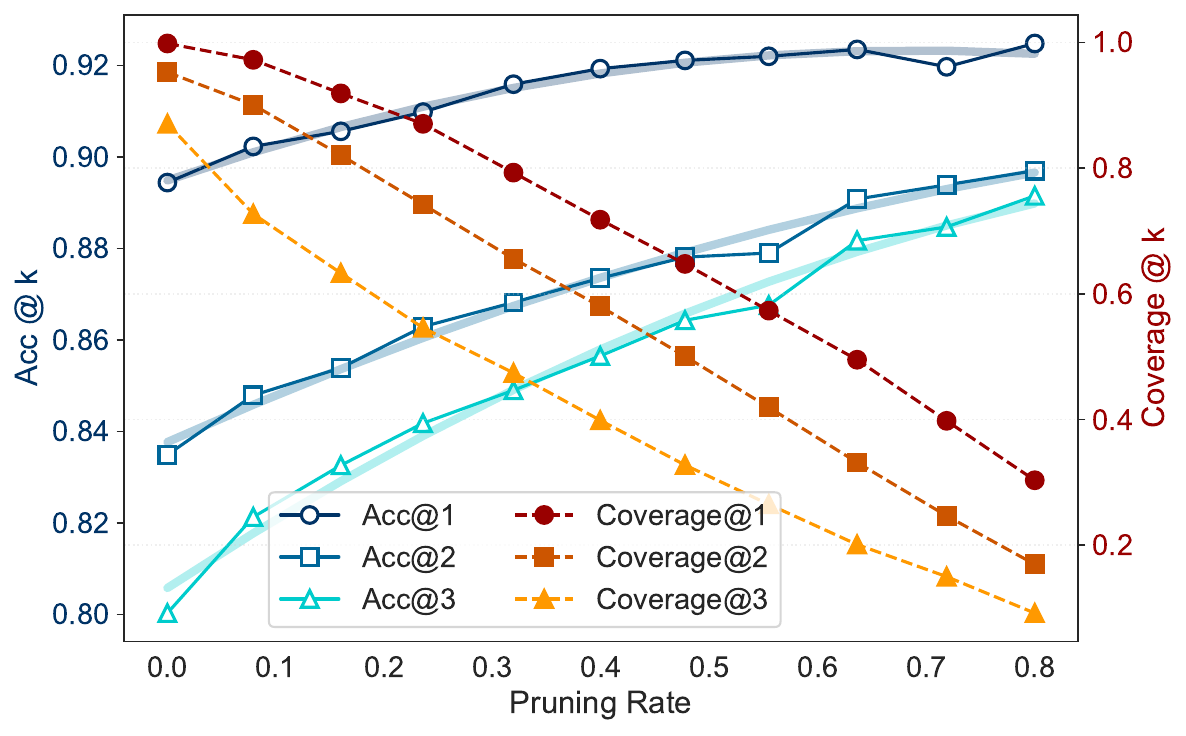}
\caption{Model Accuracy vs. Tag Coverage for Different Pruning Rates.}
\label{fig:logit}
\end{figure}




As shown in Figure~\ref{fig:logit}, our experimental results indicate that when we increase the pruning rate by setting a larger confidence threshold, the Acc@k is significantly boosted while the Coverage@k continues to decrease, which demonstrates the effectiveness of our proposed tag confidence calibration module. 
Additionally, as the pruning rate increases, the accuracy gains gradually slow down.
This characteristic allows us to set an appropriate confidence threshold in practical deployment scenarios to achieve a balance between prediction accuracy and tag coverage.

\begin{figure}[h]
  \centering
  \includegraphics[width=0.48\textwidth]{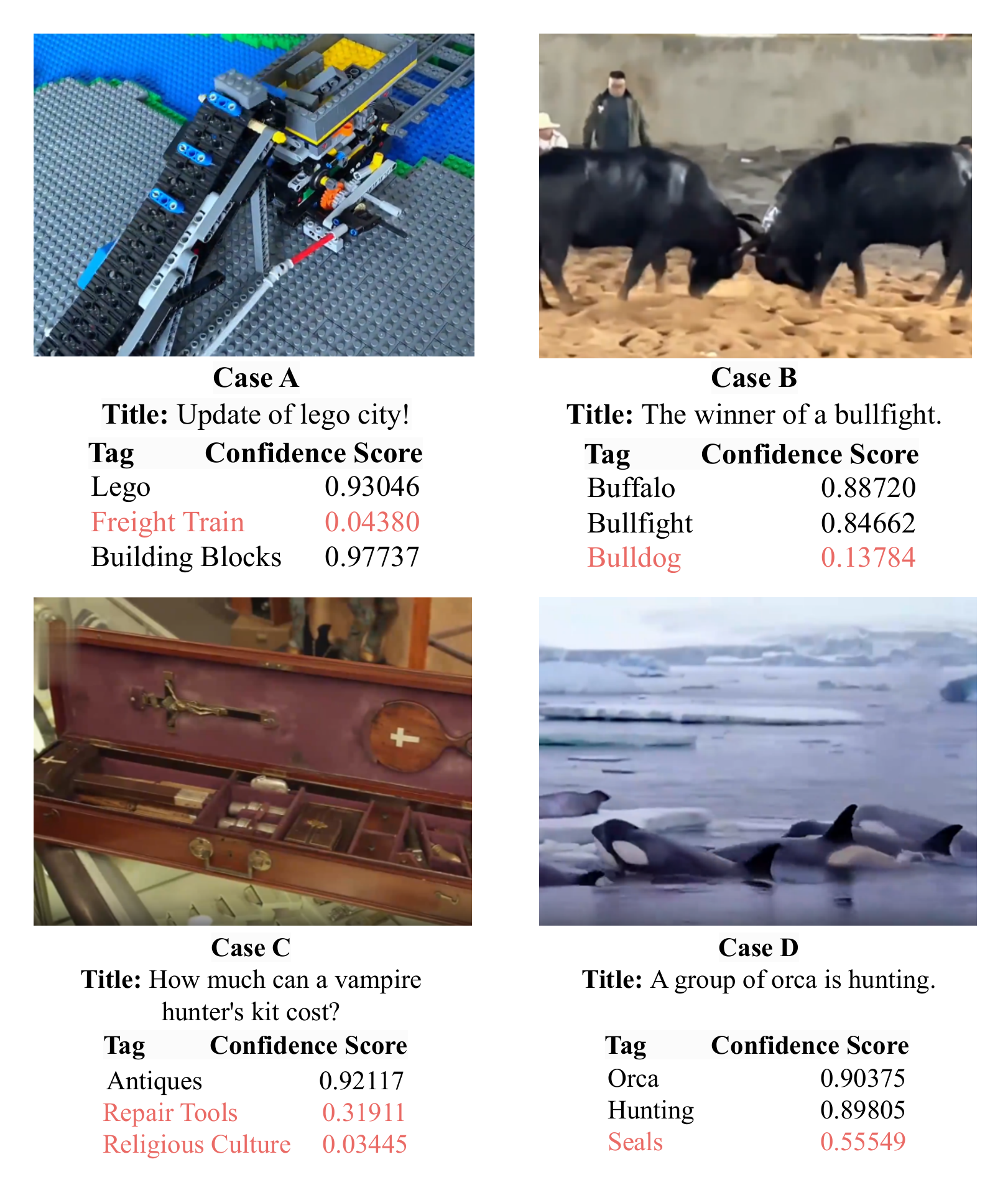}
  \caption{The online cases of tag confidence calibration module. Tags with low confidence are highlighted in red.}
  \label{fig:case_logit}
\end{figure}

Furthermore, we randomly select some cases in our deployed tagging scenario and visualize them with confidence scores in Figure~\ref{fig:case_logit}. 
We find that in Cases A, B, and C, irrelevant tags such as "Freight Train," "Bulldog," and "Religious Culture" receive low confidence scores and will be calibrated by our model, and in Case D, the weak-relevant tag, "Seals", which is a non-primary entity in the figure, receive a medium confidence score and will be ranked low in the final results, which further demonstrates the superiority of tag confidence calibration module.

\section{conclusion}
In this work, we propose an automatic tagging system based on Large Language Models (LLMs) named LLM4Tag with three key modules, characterized by completeness, continuous knowledge evolution, and quantifiability. Firstly, the graph-based tag recall module is designed to construct a small-scale relevant, comprehensive candidate tag set from a massive tag repository. Next, the knowledge-enhanced tag generation module is proposed to generate accurate tags with knowledge injection. Finally, the tag confidence calibration module is employed to generate reliable confidence tag scores. The significant improvements in offline evaluations have demonstrated its superiority and  LLM4Tag has been deployed online for content tagging.

\balance
\bibliographystyle{ACM-Reference-Format}
\bibliography{main}

\appendix

\end{document}